\documentclass[pre,superscriptaddress,draft]{revtex4}

\usepackage{multirow}

\usepackage{amssymb}
\usepackage{amsmath}
\usepackage{amsthm}
\usepackage{xcolor}

\begin{document}


\def \Z {\mathbb{Z}}
\def \R {\mathbb{R}}
\def \C {\mathbb{C}}
\def \La {\Lambda}
\def \la {\lambda}
\def \ck {l}

\newtheorem{theorem}{Theorem}
\newtheorem{lemma}{Lemma}
\newtheorem{corollary}{Corollary}

\title{Heat Current Properties of a Rotor Chain Type Model with Next-Nearest-Neighbor Interactions}

\author{Humberto C. F. Lemos}
\email[Corresponding author: ]{humbertolemos@ufsj.edu.br}
\affiliation{Departamento de Estat\'{\i}stica, F\'{\i}sica e Matem\'{a}tica, CAP - Universidade Federal de S\~ao Jo\~ao del-Rei, 36.420-000, Ouro Branco, MG, Brazil}
\author{Emmanuel Pereira}
\email{emmanuel@fisica.ufmg.br}
\affiliation{Departamento de F\'{\i}sica - Universidade Federal de Minas Gerais, CP 702, 
CEP 30.161-970, Belo Horizonte, MG, Brazil}

\begin{abstract}
In this article, to study the heat flow behavior, we perform analytical investigations in a rotor chain type model (involving inner stochastic noises) with next and next-nearest-neighbor interactions. It is known in the literature that the chain rotor model with long range interactions presents an insulating phase for the heat conductivity. But we show, in contrast with such a behavior, that the addition of a next-nearest-neighbor potential increases the thermal conductivity, at least in the low temperature regime, indicating that the insulating property is a genuine long range interaction effect. We still establish, now by numerical computations, the existence of a thermal rectification in systems with graded structures.  
\end{abstract}

\maketitle

\section{Introduction} \label{sec:intro}

 A central question in nonequilibrium statistical physics is the derivation of the macroscopic currents and their properties from the underlying microscopic models. As an example, one challenging problem that drew much attention a few decades ago was the onset of Fourier law from first principles. Fourier law states that the heat current is proportional to the gradient of temperature, i.e., to the difference of the temperatures at the ends of the system divided by its length. In a seminal work, Rieder, Lebowitz, and Lieb \cite{Rieder1967} found an anomalous heat conductivity for a chain of harmonic oscillators driven by Hamiltonian equations of motion submitted to different temperatures at the boundaries of the chain: the heat conductivity grows linearly with the system size, i.e. in other words, Fourier law does not hold, the heat current is proportional to the difference of temperature only. In Ref.\cite{Bolsterli1970}, Bolsterli, Rich, and Visscher found a normal heat conductivity (Fourier law holds) for the harmonic chain when it is under the influence of thermal reservoirs all along the chain. The temperature for the boundaries of the chain can be freely chosen, but for the inner sites, the temperatures are determined by the self-consistency condition (SC), which means that there is no net heat flow between the inner site and its linked reservoir in the steady state. With this setup, the authors showed that Fourier law holds for this model. A few decades later, the same chain of harmonic oscillators was revisited \cite{Bonetto2004}, and the question was revived. The main change is that the authors studied a $d$-dimensional system of oscillators, with $d \geq 1$. Again, all the sites of the chain are under the influence of its own thermal reservoir under SC, but now the heat baths are modeled by white noises, and so the microscopic dynamics is given by a large number of coupled stochastic ordinary differential equations. This paper triggered an avalanche of works on the subject, many of them numerical, trying to understand the necessary and/or sufficient conditions for the onset of the Fourier law. As an example, among many other microscopic models studied since then, in Refs.\cite{Gendelman2000, Giardina2000} the authors numerically studied the rotor model with nearest-neighbor (NN) nonlinear bounded interaction, finding that Fourier law holds for this one-dimensional anharmonic chain with conserved momentum, which was thought to be forbidden \cite{Prosen2000}. One of us has analytically studied a type of rotor model \cite{PereiraFalcao2006}, and found a sort of ``phase transition": Fourier law holds only at the high-temperature regime. 

Despite this approach was not able to close this Fourier law onset question, the intensive study of the heat flow on one-dimensional chains led to a more deep understanding of the subject, which allowed as a byproduct the theoretical proposal of a thermal diode \cite{Terraneo2002}: a device which conducts heat preferably in one direction, and presents a new phenomenon called thermal rectification. Again we saw a boom of works on this subject, the majority of them studied numerically, and many of them by coupling two different chains in different regimes of heat conduction, no matter if they present normal (Fourier law) or ballistic thermal conductivity. Trying to elucidate the conditions for the onset of thermal rectification, first is straightforward that the system must be inhomogeneous, but that is not sufficient: in Ref.\cite{PereiraLemosAvila} we proved the absence of thermal rectification in classical Hamiltonian harmonic chains, for any distribution for the masses along the chain, so some kind of anharmonicity is a necessary condition. In Ref.\cite{Pereira2011}, one of us established sufficient conditions for thermal rectification in general graded materials. 

Recently, the rotor model was revisited in Ref.\cite{Olivares2016}: the authors studied the rotor model with long-range (LR) attractive couplings, and they found that Fourier law holds only for sufficiently short-range interactions. In the LR regime, they found that an insulator behavior emerges, a very interesting and counter-intuitive effect. Motivated by this result, in this present paper we investigate a type of one-dimensional rotor model, but now we go beyond the NN interaction between the particles of the chain -- actually, we set up our model with a general range for the interparticle interaction potential, and we remind our analytical approach to evaluate the heat flux in section \ref{sec:model}. Using tools from stochastic calculus \cite{Oksendal}, we construct an integral formalism to evaluate the heat flow given any temperatures at the boundaries of the chain. Later, for technical reasons, we considered only a low-temperature regime for our perturbative analysis. It is worth recalling that a similar perturbative approach was proven to be rigorous in Ref.\cite{FalcaoNetoPereira2008}. In section \ref{sec:results}, we use this recently built integral formalism to evaluate heat flow for some cases. We start recapping previous known results, to assure the correctness of our results. Then we turn our attention to our model: we analytically study the linearly graded masses chain with next-nearest-neighbor (NNN) interparticle interaction. That is, we avoid the huge difficulty of the analytical investigation of rotor chain with LR interactions, but give one step in such a direction by considering a NNN potential. It is worth recalling that the investigation of the heat flow in a model with NNN interactions is interesting by itself, see, e.g., Ref.\cite{NNN}. The NN interaction coupling is always positive, while the NNN interaction coupling can be either positive or negative. In a loose way to say it, it is like we always have an attractive NN interaction between the particles, but the NNN interaction can be either attractive or repulsive. One of our goals is to find out if this model presents thermal rectification, but we also aim to investigate if a repulsive-like NNN interaction would hinder the heat flow, inspired in Ref.\cite{Olivares2016}: as we said before, they found an insulator behavior for LR attractive couplings, and this result deserves further investigation. Our analytical results show that such NNN interaction, no matter if it is attractive or repulsive, only increases the heat flow, so the insulator regime of the rotor must be a genuine LR effect, at least on the low-temperature regime. Further, we implement numerical calculations to evaluate heat flux for our NNN-interaction model, and we show that, for a graded mass chain, our system presents thermal rectification. 

The rest of this paper is organized as follows. In section \ref{sec:model} we present the model, and the used approach and derive some analytical expressions for the heat flow. In section \ref{sec:results} we describe the main results. In section \ref{sec:conclusion} we give our concluding remarks, and the Appendix is devoted to some technical notes.

\section{Model} \label{sec:model}

Let us introduce our model. We consider a chain of $N$ oscillators given by the Hamiltonian
\begin{equation} \label{eq:H}
\mathcal{H} = \sum _{j=1}^N \left[ \frac{p_j^2}{2m_j} + U^{(1)}(q_j) +
\frac{1}{2}  \sum _{\substack{ 1 \leq l \leq N; \\ l \neq j}}  U^{(2)}(q_j-q_l) \right],
\end{equation}
where $q_{j}$ and $p_{j}$ give us, respectively, position and momentum for $j$-th particle of the chain, $m_{j}$ is particle mass, and it is pinned to its equilibrium position $q_{j} = 0$ by a harmonic interaction $U^{(1)}(q_j) = M_j q_j^2/2$, henceforth named on-site potential. The particles interact with each other by a bounded anharmonic interparticle potential
\begin{equation} \label{eq:U2}
U^{(2)}(q_j-q_l) = \lambda _{j,l} [1 - \cos ( \kappa(q_j - q_l) ) ],
\end{equation}
where $\lambda _{j,l}$ is the coupling strength, and $\kappa$ is a parameter usually taken as 1 in the other studies of the rotor model. In other words, we study heat flux on a version of a well-known rotor model \cite{PereiraFalcao2006}. Definition \eqref{eq:U2} above is quite general, but in this work, we take only symmetric interaction coupling $\lambda _{j,l} = \lambda _{l,j}$. It is worth noticing that the Hamiltonian \eqref{eq:H} poses no restriction on the range of the interparticle interaction, and we can both study nearest-neighbor (NN) or long-range (LR) models, among others. The dynamics is given by Hamilton equations of motion coupled to stochastic white noises which mimic the contact of the system with thermal reservoirs (at least for the noise at the boundaries, details ahead), namely
\begin{subequations} \label{eq:dyn}
\begin{align}
\label{eq:dynq}
dq_j &= \frac{\partial \mathcal{H}}{\partial p_j} \, dt = \frac{p_j}{m_j} \, dt ,
\\
dp_j &= - \frac{\partial \mathcal{H}}{\partial p_j} - \zeta _j p_j dt + \gamma _j^{1/2} dB_j = - M_j q_j dt - \sum _{l \neq j} U^{\prime(2)} \, dt - \zeta _j p_j dt + \gamma _j^{1/2} dB_j , \label{eq:dynp}
\end{align}
\end{subequations}
where prime denotes the derivative with respect to $q_j$, viz.
\begin{equation} \label{eq:U2prime}
U^{\prime(2)} (q_{j} - q_{l}) = \lambda _{j,l} \, \kappa \sin (\kappa(q_j - q_l)) = U^{\prime (2)}_{j,l},
\end{equation}
where the last equality above is just a definition for the shortcut notation $U^{\prime (2)}_{j,l}$. On Eq. \eqref{eq:dynp}, each $dB_{j}$ is a zero mean independent Wiener process, i.e.
\begin{equation} \label{eq:dB}
    \left\langle dB_{j} (t) \right\rangle = 0 , \quad \left\langle dB_{j} (t) dB_{j'} (t') \right\rangle = \delta_{j,j'} \delta(t-t') dt,
\end{equation}
for any given sites $j,j'$ of the chain and times $t,t' > 0$. We also have $\gamma_{j} = 2 m_{j} \zeta_{j} T_{j}$, where $\zeta_{j}$ is heat bath coupling constant for $j$-th site, and $T_{j}$ is the temperature of the $j$-th heat bath.

From now on, for the sake of understanding, we recall the main steps of our approach. Further details can be found in previous works \cite{PereiraFalcao2004, PereiraFalcao2006, PereiraLemosAvila}. Symmetrically defining the energy $\mathcal{H}_{j}$ for the $j$-th particle as $\mathcal{H} =\sum_{j} \mathcal{H}_{j}$, we get
\begin{equation} \label{eq:Hj}
\mathcal{H}_{j} = \frac{p_j^2}{2m_j} + \frac{1}{2} \, M_j q_j^2 +
\frac{1}{2}  \sum _{ l \neq j }  U^{(2)}(q_j-q_l).
\end{equation}
Using mathematical tools from It\^{o} stochastic calculus \cite{Oksendal}, we can obtain 
\begin{equation} \label{eq:meanHj}
\left\langle \frac{d\mathcal{H}_j}{dt} \right\rangle = \left\langle \mathcal{F}_{\rightarrow j} \right\rangle - \left\langle \mathcal{F}_{j\rightarrow} \right\rangle + \left\langle R_j \right\rangle,
\end{equation}
where $\left\langle \cdot \right\rangle$ denotes expectation with respect to white noise distribution, and
\begin{subequations}
\begin{align}
\label{eq:Rj}
R_j &= \zeta _j \left( T_j - \frac{p_j^2}{m_j} \right) \, ,
\\
\label{eq:F2j}
\mathcal{F}_{\rightarrow j} &= \frac{1}{2} \sum _{l < j} U^{\prime(2)} (q_l - q_j) \left(\frac{p_j}{m_j} + \frac{p_l}{m_l}\right) \, ,
\\
\label{eq:Fj2}
\mathcal{F}_{j\rightarrow} &= \frac{1}{2} \sum _{l > j} U^{\prime(2)} (q_j - q_l) \left(\frac{p_j}{m_j} + \frac{p_l}{m_l}\right)  \, .
\end{align}
\end{subequations}
Detailing, $R_{j}$ tells us about the average energy exchange between the $j$-th site and its thermal reservoir, while $\mathcal{F}_{\rightarrow j} (\mathcal{F}_{j\rightarrow})$ gives us the energy flux from (to) $l$-th sites to (from) $j$-th site; in other words, the heat flux inside the chain.

We aim to study heat flux on the nonequilibrium stationary state (NESS), so we take $T_{1} \neq T_{N}$ for temperatures at the boundaries of the chain. For inner sites, $T_{j}$ will be given by self-consistency condition, which means that on NESS there will be, on average, no energy exchange between $j$-th site of chain and its bath, i.e. $\left\langle R_j \right\rangle = 0$. In other words, the inner stochastic reservoirs are not real thermal baths, they only represent some phonon scattering process given by interactions not directly presented in the Hamiltonian. Since NESS is characterized by stationary energy flux, we have
\begin{equation} \label{eq:NESS}
\left\langle \frac{d\mathcal{H}_j}{dt} \right\rangle = 0,
\end{equation}
and therefore $\left\langle \mathcal{F}_{\rightarrow j} \right\rangle = \left\langle \mathcal{F}_{j\rightarrow} \right\rangle$, for any $2 \leq j \leq N-1$. In other words, if for example, we have $T_{1} >  T_{N}$, the thermal reservoir connected to the left site injects energy on the chain, the energy flows through it and leaves it on the right boundary. Hence, to know heat flux on NESS, we must evaluate $\left\langle \mathcal{F}_{\rightarrow j} \right\rangle$ or $\left\langle \mathcal{F}_{j\rightarrow} \right\rangle$ for any inner site $j$.

Aiming to solve stochastic ODE's \eqref{eq:dyn}, we now define phase space vector $\varphi = (q,p)^{\dagger} \in \R^{2N}$, i.e. $\varphi_{j} = q_{j}$ and $\varphi_{j+N} = p_{j}$, for any $1 \leq j \leq N$. We rewrite dynamics \eqref{eq:dyn} as
\begin{equation} \label{eq:dynphase}
d\varphi = - A \varphi dt - \mathcal{U}^{\prime}(\varphi) dt + \sigma dB,
\end{equation}
where $A$ and $\sigma$ are $2N \times 2N$ matrices respectively given by
\begin{equation} \label{eq:A}
A = 
 \begin{pmatrix}
  0 & -m^{-1} \\
  M & \zeta
 \end{pmatrix},
\quad
\sigma = 
 \begin{pmatrix}
  0 & 0 \\
  0 & \sqrt{2m \zeta T}
 \end{pmatrix}.
\end{equation}
In equation above, both matrices are described in four $N \times N$ blocks, and despite reduntant notation, $m$ means the diagonal matrix for the masses, $m_{j,l} = m_{j} \delta_{j,l}$, and the same holds for $N \times N$ diagonal matrices $M$, $\zeta$ and $T$; and the nonlinear term $\mathcal{U}^{\prime}$ in Eq. \eqref{eq:dynphase} reminds us about $U^{(2)}$ derivative with respect to $q$ -- note that $\mathcal{U}^{\prime}$ is nonzero only for indices $j>N$. Also, again using a redundant notation, $dB$ is a $2N$-vector whose components are $dB_{j}=0$, and $dB_{j+N}$ is the white noise acting on the $j$-th site of the chain -- see Eq. \eqref{eq:dynp} -- for any $1 \leq j \leq N$. To obtain the heat flux on NESS, we fix any site $\alpha$ in the bulk of the chain and evaluate $\left\langle \mathcal{F}_{\alpha \rightarrow} \right\rangle$ given by Eq.\eqref{eq:Fj2}, which will be defined below as
\begin{widetext}
\begin{equation} \label{eq:Omega}
\left\langle \Omega (\varphi) \right\rangle = \lim_{t \to \infty} \left\langle \mathcal{F}_{\alpha \rightarrow} (\varphi(t)) \right\rangle = \lim_{t \to \infty} \frac{1}{2} \sum _{\beta > \alpha} \lambda _{\alpha,\beta} \kappa \left\langle \sin \big( \kappa \big( \varphi _{\alpha}(t) - \varphi _{\beta}(t) \big) \big) \left(\frac{\varphi _{\alpha+N}(t)}{m_{\alpha}} + \frac{\varphi _{\beta+N}(t)}{m_{\beta}}\right) \right\rangle ,
\end{equation}
\end{widetext}
where we have used $U^{\prime(2)}$ given by Eq.\eqref{eq:U2prime}. We emphasize that the average of $\Omega (\varphi)$ defined above gives us the heat flux on NESS, and our main goal is to evaluate it. But as we can see from Eq.\eqref{eq:dyn}, the equations of motion for this system are a set of $2N$ first-order coupled nonlinear stochastic ODEs, and to find a solution for such a set of equations is a really hard, if not impossible, task. We then proceed as follows: first, we find the solution for a simplified process denoted as $\phi$, which is related to the complete one, named $\varphi$. This easier problem is obtained by taking interparticle coupling as identically zero, i.e. $\lambda_{j,l}=0$. So now we have $2N$ linear decoupled stochastic ODEs, written as
\begin{equation} \label{eq:dynphase0}
d\phi = - A \phi \, dt + \sigma dB.
\end{equation}
The solution for Eq. \eqref{eq:dynphase0} is the well known Ornstein-Uhlenbeck process
\begin{equation} \label{eq:phit}
\phi (t) = e^{-tA} \phi (0) + \int _0^t e^{-(t-s) A}\sigma dB (s).
\end{equation}
Defining $\left\langle \cdot \right\rangle_{0}$ as the average over noises realisations for simplified process \eqref{eq:dynphase0}, we have
$$
\left\langle \phi (t) \right\rangle_{0} = e^{-tA} \left\langle \phi (0) \right\rangle_{0} ,
$$
where we have used an important property from It\^{o} stochastic calculus that guarantees that
$$
\left\langle \int _{S}^{T} \psi(s) dB (s) \right\rangle_{0} = 0,
$$
for some class of well behaved functions $\psi$, details in Ref.\citep{Oksendal}. Since $A$ is a stable matrix \cite{Snyders}, we have $e^{-tA} \phi (0) \to 0$ as $t \to +\infty$, for any given initial condition, so without loss of generality we take $\phi (0) = 0$. Then \eqref{eq:dynphase0} is a zero mean Gaussian process, whose covariance is
\begin{equation} \label{eq:covphi}
\left\langle \phi (t) \phi ^{\dagger}(t') \right\rangle_{0} = \mathcal{C} (t,t'),
\end{equation}
where
\begin{equation} \label{eq:Cttlinha}
\mathcal{C} (t,t') = 
  \begin{cases}
    e^{-(t-t') A} \mathcal{C} (t',t') & , \text{ if } t \geq t' \\
    \mathcal{C} (t,t) e^{-(t'-t) A^{\dagger}} & , \text{ if } t \leq t' ,
  \end{cases}
\end{equation}
with
\begin{equation} \label{eq:Ctt}
\mathcal{C} (t,t) = \int _0^t ds \, e^{-sA} \sigma ^2 e^{-sA^{\dagger}}.
\end{equation}  
From a straightforward computation, it follows that, for a single site $j$
\begin{equation} \label{eq:expAt}
e^{-tA_{(j)}} = e^{-\frac{\zeta _{j}}{2} \, t} \left( \cosh (\rho _{j} t) I_{2} + \frac{\sinh (\rho _{j} t)}{\rho _{j}} \, B_{(j)} \right),
\end{equation}
where $\rho_{j} = [(\zeta_{j}/2)^{2} - M_{j}/m_{j}]^{1/2}$ and $A_{(j)}$ is the $2 \times 2$ matrix related to $A$ for a single site $j$, $I_{2}$ is the identity matrix and
$$
B_{(j)} = \begin{pmatrix}
   \frac{\zeta _{j}}{2}  & m_{j}^{-1} \\
   - M_{j} & - \frac{\zeta _{j}}{2}
 \end{pmatrix}.
$$
Evaluating Eq. \eqref{eq:Ctt} for $t \to +\infty$, we get NESS covariance for isolated process $\phi$
\begin{equation} \label{eq:C}
C = \int _{0}^{\infty} ds \, e^{-sA} \sigma ^2 e^{-sA^{\dagger}} = \begin{pmatrix}
   M^{-1} T  & 0 \\
   0 & mT
 \end{pmatrix},
\end{equation}
and we can see that the covariance $C$ is a diagonal matrix for the simplified process $\phi$. As a final remark for covariance, if $t$ and $t'$ are sufficient large, we can approach Eq.\eqref{eq:Cttlinha} as
\begin{equation} \label{eq:Cttlinhaapprox}
\mathcal{C} (t,t') = 
  \begin{cases}
    e^{-(t-t') A} C + \mathcal{O} \left( e^{-(t+t') \zeta} \right) & , \text{ if } t \geq t' \\
    C e^{-(t'-t) A^{\dagger}} + \mathcal{O} \left( e^{-(t+t') \zeta} \right) & , \text{ if } t \leq t' .
  \end{cases}
\end{equation}

To recover the effects of anharmonic interparticle potential $U^{(2)}$ on the system, we use the Girsanov theorem \cite{Oksendal}, which says that to evaluate the average for any quantity $f$ that depends on the complete process $\varphi$, we can compute the average for the same quantity $f$ depending on simplified process $\phi$, corrected by a factor $Z(t)$
$$
\left\langle f(\varphi(t)) \right\rangle = \left\langle f(\phi(t)) Z(t) \right\rangle_{0},
$$
which is given by
\begin{equation} \label{eq:Zt}
Z(t) = \exp \left( \int _0^t u \cdot dB(s) - \frac{1}{2} \int _0^t \|u\|^2 ds \right) \, , 
\end{equation}
where $u \in \R^{2N}$ is related to the difference between complete and simplified processes. Namely, for any index $1 \leq j \leq N$, we have
\begin{eqnarray} \label{eq:u}
    u_{j} &=& 0  \\
    \gamma _{j}^{1/2} u _{j+N} &=& \sum_{l \neq j} U^{\prime(2)}_{j,l} = \sum_{l \neq j} \lambda_{j,l} \kappa \sin (\kappa (\phi_{j} - \phi_{l})). \nonumber
\end{eqnarray}
After some tedious but straightforward calculations, we find
\begin{equation} \label{eq:Ztrot}
Z(t) = \exp \left[ - \Delta F(\phi(t)) 
- \int _0^t W(\phi(s)) \, ds \right] \, , 
\end{equation}
where $\Delta F(\phi(t)) = F(\phi(t)) - F(\phi(0))$, with
\begin{equation}
\label{eq:Ftrot}
F(\phi(t)) = \frac{1}{2\zeta _j m_j T_j} \left( \sum _{l \neq j} \lambda_{j,l} \kappa \sin \big( \kappa \big( \phi_{j}(t) - \phi_{l}(t) \big) \big) \right) \phi _{j+N}(t),
\end{equation}
and $W(\phi(s)) = W_{1}(\phi(s)) + W_{2}(\phi(s)) + W_{3}(\phi(s)) + W_{4}(\phi(s))$, with
\begin{subequations} \label{eq:Wrot}
\begin{align}
\label{eq:W1rot}
W_{1}(\phi(s)) &= \sum_{j} \sum _{l \neq j} \frac{\lambda_{j,l} \kappa M_j \phi _{j}(s)}{2\zeta _j m_j T_j} \, \sin \big( \kappa \big( \phi_{j}(s) - \phi_{l}(s) \big) \big)   ,
\\
\label{eq:W2rot}
W_{2}(\phi(s)) &= \sum_{j} \sum _{l \neq j} \frac{\lambda_{j,l} \kappa \zeta _j \phi _{j+N}(s) }{2 \zeta _j m_j T_j} \, \sin \big( \kappa \big( \phi_{j}(s) - \phi_{l}(s) \big) \big) ,
\\
\label{eq:W3rot}
W_{3}(\phi(s)) &= - \sum_{j} \sum _{l \neq j} \frac{\lambda_{j,l} \kappa^{2} \phi _{j+N}(s)}{2 \zeta _j m_j T_j} \, \cos \big( \kappa \big( \phi_{j}(s) - \phi_{l}(s) \big) \big) \ \left( \frac{\phi _{j+N}(s)}{m_j} - \frac{\phi _{l+N}(s)}{m_l}\right) ,
\\
\label{eq:W4rot}
W_{4}(\phi(s)) &= \sum_{j} \sum _{l,l' \neq j} \frac{\lambda_{j,l} \lambda_{j,l'} \kappa^{2}}{4 \zeta _j m_j T_j} \, \sin \big( \kappa \big( \phi_{j}(s) - \phi_{l}(s) \big) \big) \sin \big( \kappa \big( \phi_{j}(s) - \phi_{l'}(s) \big) \big).
\end{align}
\end{subequations}

We now develop a perturbative approach for our calculations, taking nonlinear coupling $\lambda$ as a small perturbative parameter. We note from Eq.\eqref{eq:W4rot} that $W_{4}$ depends on $\lambda^{2}$, and so this term will be dropped on a first-order expansion. A first-order expansion on $\lambda$ gives us
\begin{equation} 
\left\langle \Omega (\varphi) \right\rangle = \frac{\left\langle  \Omega(\phi) e^{-\Delta F - \int W ds}\right\rangle_0}{\left\langle e^{-\Delta F - \int W ds}\right\rangle_0} = \left\langle \Omega(\phi)\right\rangle_0  - \left\langle \Omega(\phi); \Delta F \right\rangle_0 - \left\langle \Omega(\phi); \int W ds \right\rangle_0 +  \mathcal{O} \left( \lambda^3\right),
\label{eq:perturbrot}
\end{equation}
where the semicolon means truncated expectation value given by
$$
\left\langle f;g \right\rangle = \left\langle fg \right\rangle - \left\langle  f \right\rangle \left\langle g \right\rangle .
$$
We remind our definition \eqref{eq:Omega} and emphasize that all averages above must be taken on limit $t \to \infty$. It may be confusing to see on Eq.\eqref{eq:perturbrot} an expression up to order $\mathcal{O} \left( \lambda^3\right)$: despite we have taken only first-order terms in our perturbative parameter, we already have a $\lambda$ on $\Omega$ definition, as one can see on Eq.\eqref{eq:Omega}. This will be clear after we evaluate our first term on expression -- see, e.g. Eq.\eqref{eq:Ftdone}.

To obtain the heat flux we must now evaluate each term on Eq.\eqref{eq:perturbrot}. It is easy to see that $\left\langle \Omega(\phi)\right\rangle_0 = 0$. Indeed, Eq. \eqref{eq:Omega} shows that it depends on $C_{k,k'+N} = 0$, as we can see on Eq. \eqref{eq:C}. For a similar reason we get $\left\langle \Omega(\phi) F(\phi(0)) \right\rangle_0 = 0$. As an example of a non-vanishing average, we show the main steps in evaluation for 
\begin{widetext}
\begin{eqnarray}
\left\langle \Omega; F_t \right\rangle _{0} &:=& \lim _{t \to +\infty} \left\langle \Omega(\phi(t)); F(t) \right\rangle _{0} = 
   \lim _{t \to +\infty} \sum _{\beta > \alpha ; j; l \neq j} 
  \frac{\lambda_{\alpha,\beta} \lambda_{j,l} \kappa }{4 \zeta_j m_j T_j} \times
  \nonumber \\
  && \times \left\langle  \sin \big( \kappa \big( \phi _{\alpha}(t) - \phi _{\beta}(t) \big) \big) \left(\frac{\phi _{\alpha+N}(t)}{m_{\alpha}} + \frac{\phi _{\beta+N}(t)}{m_{\beta}}\right); \sin \big( \kappa \big( \phi _j(t) - \phi _l(t) \big) \big) \phi _{j+N}(t) \right\rangle _{0} . \label{eq:Ftgen}
\end{eqnarray}
\end{widetext}
To deal with such expressions, we write sine functions as complex exponentials, i.e. $\sin (\kappa (\phi_{\alpha} - \phi _{\beta})) = (e^{+i \kappa (\phi _{\alpha} - \phi _{\beta})} - e^{-i \kappa (\phi _{\alpha} - \phi _{\beta})})/2i$. And since our average is over a Gaussian measure, we use the following approach to evaluate such quantities. Since 
$$
\left\langle \cdot \right\rangle_{0} = \mathcal{N}^{-1} \int \cdot \ e^{-\frac{1}{2} (\phi, \mathcal{C}^{-1} \phi)} d\phi =  \mathcal{N}^{-1} \int \cdot \ e^{-\frac{1}{2} (\phi, \mathcal{C}^{-1} \phi)} e^{i \kappa (h, \phi)} d\phi \bigg|_{h=0} = G(h)\bigg|_{h=0},
$$
where $\mathcal{N}$ is a normalization factor, and $(\phi, \mathcal{C}^{-1} \phi)$ is the canonical inner product on $\R^{2N}$. On the last equation, we have defined an auxiliary function $G(h)$, where $h \in \R^{2N}$ is an arbitrary vector which, for the quantity above, is taken as zero after we evaluate the integral. This procedure can also help us to evaluate other quantities, for example
\begin{equation}
\left\langle \phi_{j+N}(t) e^{+i \phi _{\alpha}(t)} \right\rangle_{0} =  \frac{1}{i \kappa} \frac{\partial}{\partial h_{j+N}}  G(h)\bigg|_{h_{\alpha}=1} = \mathcal{C}_{\alpha , j + N} (t,t) \, e^{-\frac{1}{2} \mathcal{C}_{\alpha , \alpha} (t,t)}, \label{eq:derivative}
\end{equation}
where $h_{\alpha}=1$ is taken after evaluate derivative to keep a remaining $\phi_{\alpha}$ on imaginary exponential, all other components of vector $h$ are taken as zero. By choosing properly the derivatives and non-zero components, we can show that
\begin{widetext}
\begin{eqnarray}
\left\langle \Omega; F_t \right\rangle _{0} &=& \sum _{\beta > \alpha} \sum _{l \neq \alpha} 
\frac{\lambda _{\alpha,\beta} \lambda _{\alpha,l}} {8\zeta _{\alpha} m_{\alpha} \kappa}
e^{-\frac{1}{2} \big( C_{\beta,\beta} + C_{l,l} \big)}
\left(
e^{-\big( C_{\beta,l} + 2 C_{\alpha,\alpha} \big)} - e^{+C_{\beta,l}}
\right) +
\nonumber
\\
& & + \sum _{\beta > \alpha} \sum _{l \neq \beta} 
\frac{\lambda _{\alpha,\beta} \lambda _{\beta,l}} {8\zeta _{\beta} m_{\beta} \kappa}
e^{-\frac{1}{2} \big( C_{\alpha,\alpha} + C_{l,l} \big)}
\left( e^{+ C_{\alpha,l}} - e^{-\big( C_{\alpha,l} + 2 C_{\beta,\beta} \big)}
\right) . \label{eq:Ftdone}
\end{eqnarray}
\end{widetext}

Equation \eqref{eq:Ftdone} can be evaluated for any regime of temperatures, but it does not tell us much in this form. We, from now on, develop an approach for studying heat flux in a low-temperature regime, i.e. when $T_j$ is small for any site on the chain. Here, a small temperature means that $T_{j} < 1$, we give more details in appendix \ref{app:dimless} ahead. We can see from equations \eqref{eq:covphi}-\eqref{eq:C} that the covariance $\mathcal{C}$ is proportional to the temperature, so from the leading term of Taylor series for exponentials on Eq. \eqref{eq:Ftdone} we get
\begin{equation} \label{eq:FtTpeq}
- \left\langle \Omega; F_t \right\rangle_{0} = \sum _{\beta > \alpha} \frac{\lambda _{\alpha,\beta}^2}{4 \kappa}
\left[ 
\left(
\frac{T_{\alpha}}{M_{\alpha}} + \frac{T_{\beta}}{M_{\beta}}
\right)
\left(
\frac{1}{\zeta _{\beta} m_{\beta}} - \frac{1}{\zeta _{\alpha} m_{\alpha}}
\right)
\right] +
\sum _{\beta > \alpha} \sum _{l \neq \alpha , \beta} \frac{\lambda _{\alpha,\beta}}{4 \kappa} 
\left[ 
\frac{\lambda _{\beta,l}}{\zeta _{\beta} m_{\beta}} \frac{T_{\beta}}{M_{\beta}}
-
\frac{\lambda _{\alpha,l}}{\zeta _{\alpha} m_{\alpha}} \frac{T_{\alpha}}{M_{\alpha}}
\right] .
\end{equation}
A first glance at Eq. \eqref{eq:FtTpeq} may be deceptive and lead someone to believe that we have a first-order approach on covariance $\mathcal{C}$, but a further look at Eq. \eqref{eq:Ftgen} show us that we had a $T_{j}^{-1}$ from the start. So actually our leading term is of order $\mathcal{O} \left( \mathcal{C}^2 \right)$, and it will be the leading term as we use the same approach to handle the remaining terms. For example, for $W_{1}$ given in \eqref{eq:W1rot}, we have
\begin{widetext}
\begin{eqnarray*}
- \left\langle \Omega; W_1 \right\rangle_{0} &=&
- \lim _{t \to +\infty} \left\langle \Omega(\phi(t)); \int _0^t W_{1}(\phi(s)) \, ds \right\rangle_{0} =
\\
&=& - \frac{1}{2} \lim _{t \to +\infty} \sum _{\beta > \alpha} \sum _{j} \sum _{l \neq j}
 \frac{\lambda _{\alpha,\beta} \lambda _{j,l} \kappa M_j}{2\zeta _j m_j T_j} \times
\\
& &\times \int _0^t ds \left\langle
\sin \Big(\phi _{\alpha}(t) - \phi _{\beta}(t) \Big)
\bigg( \frac{\phi _{\alpha+N}(t)}{m_{\alpha}} +
 \frac{\phi _{\beta+N}(t)}{m_{\beta}} \bigg);
\sin \Big(\phi _j(s) - \phi _l(s)\Big)  \phi _{j}(s)
\right\rangle_{0} .
\end{eqnarray*}
\end{widetext}
Calculations are extensive from now on. We again use the auxiliary function $G(h)$ approach, as we did on \eqref{eq:derivative}, but now we will come up with a second-order derivative on $h$. It will raise many terms, but they are all like
$$
\lim _{t \to +\infty} \sum _{\beta , j, l}
\frac{\lambda _{\alpha,\beta} \lambda _{j,l} M_j}{8 \zeta _j m_{\alpha} m_j \kappa T_j} \int _0^t ds \ \mathcal{C}_{\alpha+N,j} (t,s) e^{-\frac{1}{2} (h,\mathcal{C}h) } \bigg|_{h_{1}-h_{2}} ,
$$
or like
$$
\lim _{t \to +\infty} \sum _{\beta , j, l}
\frac{\lambda _{\alpha,\beta} \lambda _{j,l} M_j}{8 \zeta _j m_{\alpha} m_j \kappa T_j} \int _0^t ds \ \mathcal{C}_{\alpha+N,j} (t,s) \mathcal{C}_{\alpha,j} (t,s)
e^{-\frac{1}{2} (h,\mathcal{C}h) } \bigg|_{h_{1}+h_{2}},
$$
where $h_1$ or $h_2$ refer to the signs that came from imaginary exponentials that define sine functions. Namely, for $h_{1}$ we take $h_{\alpha}=+1$, $h_{\beta}=-1$, $h_{j}=+1$ and $h_{l}=-1$, while for $h_{2}$ we only change to $h_{j}=-1$ and $h_{l}=+1$. To deal with those integrals on $ds$, we use approximation presented on Eq. \eqref{eq:Cttlinhaapprox}, and analytically calculate them. Calculations are tedious but straightforward, and after them we obtain
\begin{eqnarray} \label{eq:W1rot_Tpeq_calc}
- \left\langle \Omega; W_{1} \right\rangle_{0} &=& \sum_{\beta > \alpha} \left( \frac{\lambda_{\alpha,\beta}^2}{4 m_{\alpha} M_{\alpha} \kappa} \, \frac{T_{\alpha}}{\zeta_{\alpha}} - \frac{\lambda_{\alpha,\beta} \lambda_{\beta,\alpha}}{4 m_{\beta} M_{\beta} \kappa} \, \frac{T_{\beta}}{\zeta_{\beta}} \right) + \sum_{\beta > \alpha} \sum_{l \neq \alpha,\beta} \left( \frac{\lambda_{\alpha,\beta} \lambda_{\alpha,l}}{4 m_{\alpha} M_{\alpha} \kappa} \, \frac{T_{\alpha}}{\zeta_{\alpha}} - \frac{\lambda_{\alpha,\beta} \lambda_{\beta,l}}{4 m_{\beta} M_{\beta} \kappa} \, \frac{T_{\beta}}{\zeta_{\beta}} \right) + 
\\
&+& \sum_{\beta > \alpha} \frac{\lambda_{\alpha,\beta} }{4 m_{\alpha} m_{\beta} \kappa} \, \frac{\zeta_{\alpha} + \zeta_{\beta}}{D_{\alpha,\beta}} \,  \left( \lambda_{\beta,\alpha} T_{\alpha} - \lambda_{\alpha,\beta} T_{\beta} \right) + \sum_{\beta > \alpha} \frac{\lambda_{\alpha,\beta}}{4 m_{\alpha} m_{\beta} \kappa D_{\alpha,\beta}} \, \left( \frac{M_{\alpha}}{m_{\alpha}} - \frac{M_{\beta}}{m_{\beta}}  \right) \left( \frac{\lambda_{\beta,\alpha} T_{\alpha}}{\zeta_{\beta}} + \frac{\lambda_{\alpha,\beta} T_{\beta}}{\zeta_{\alpha} D_{\alpha,\beta}} \right) +
\nonumber
\\
&+& \sum_{\beta > \alpha} \left( \frac{\lambda_{\alpha,\beta}^2 M_{\alpha}}{4 m_{\alpha}^2 M_{\beta} \kappa} \, \frac{\zeta_{\beta} (\zeta_{\alpha} + \zeta_{\beta})}{\zeta_{\alpha} D_{\alpha,\beta}} \, T_{\beta} - \frac{\lambda_{\alpha,\beta} \lambda_{\beta,\alpha} M_{\beta}}{4 m_{\beta}^2 M_{\alpha} \kappa} \, \frac{\zeta_{\alpha} (\zeta_{\alpha} + \zeta_{\beta})}{\zeta_{\beta} D_{\alpha,\beta}} \, T_{\alpha} \right)  +
\nonumber
\\
&+& \sum_{\beta > \alpha} \left( \frac{\lambda_{\alpha,\beta}^2 M_{\alpha}}{4 m_{\alpha}^2 M_{\beta} \kappa} \, \left( \frac{M_{\alpha}}{m_{\alpha}} - \frac{M_{\beta}}{m_{\beta}} \right)  \frac{T_{\beta}}{\zeta_{\alpha} D_{\alpha,\beta}} + \frac{\lambda_{\alpha,\beta} \lambda_{\beta,\alpha} M_{\beta}}{4 m_{\beta}^2 M_{\alpha} \kappa} \, \left( \frac{M_{\alpha}}{m_{\alpha}} - \frac{M_{\beta}}{m_{\beta}} \right) \frac{T_{\alpha}}{\zeta_{\beta} D_{\alpha,\beta}} \right),
\nonumber
\end{eqnarray}
where 
\begin{equation} \label{eq:Djl}
D_{\alpha,\beta} = (\zeta _{\alpha} + \zeta _{\beta}) \left( \zeta _{\beta} \frac{M_{\alpha}}{m_{\alpha}} + \zeta _{\alpha} \frac{M_{\beta}}{m_{\beta}} \right) + \left( \frac{M_{\alpha}}{m_{\alpha}} - \frac{M_{\beta}}{m_{\beta}} \right)^2.
\end{equation}
Following the same approach for the remaining terms, we get
\begin{equation} \label{eq:W2rotTpeq}
- \left\langle \Omega; W_{2} \right\rangle_{0} = \sum_{\beta > \alpha} \frac{\lambda_{\alpha,\beta} }{2 m_{\alpha} m_{\beta} \kappa} \, \frac{\zeta_{\alpha} + \zeta_{\beta}}{D_{\alpha,\beta}} \, \big( \lambda_{\beta,\alpha} T_{\alpha} - \lambda_{\alpha,\beta} T_{\beta} \big) ,
\end{equation}
and
\begin{equation} \label{eq:W3rotorTpeq}
- \left\langle \Omega; W_{3} \right\rangle_{0} = - \sum _{\beta > \alpha} \frac{\lambda_{\alpha,\beta} }{2 m_{\alpha} m_{\beta} \kappa}\left( \frac{M_{\alpha}}{m_{\alpha}} - \frac{M_{\beta}}{m_{\beta}} \right) \frac{1}{D_{\alpha,\beta} } \left( \frac{\lambda_{\beta,\alpha} T_{\alpha} }{\zeta_{\beta} } + \frac{\lambda_{\alpha,\beta} T_{\beta} }{\zeta_{\alpha} }\right)
\end{equation}
Summarizing, adding each term \eqref{eq:FtTpeq}-\eqref{eq:W3rotorTpeq} for the flux \eqref{eq:perturbrot}, we get
\begin{equation} \label{eq:fluxTpeq}
    \mathcal{F}_{\alpha \rightarrow} = - \left\langle \Omega; F_t \right\rangle_{0} - \left\langle \Omega; W_{1} \right\rangle_{0} - \left\langle \Omega; W_{2} \right\rangle_{0} - \left\langle \Omega; W_{3} \right\rangle_{0} .
\end{equation}
It is not worth obtaining a closed expression for \eqref{eq:fluxTpeq} right now, it is better to do it for each case in the next section.

\section{Results} \label{sec:results}

\subsection{Short reminder of previous results} \label{ssec:previous}

We start checking results \eqref{eq:FtTpeq}-\eqref{eq:W3rotorTpeq} on previously studied models. Initially, we consider the homogeneous chain, i.e. mass $m_{j}=m$, on-site harmonic potential $M_{j}=M$, and bath coupling to the chain $\zeta_{j}=\zeta$ are the same for any site $1 \leq j \leq N$. We also take only homogeneous NN interactions given by
$$
\lambda_{j,l} = \begin{cases}
\lambda > 0 & \mbox{, if } |j-l| = 1, \\
0 & \mbox{, otherwise}.
\end{cases}
$$
In such case, for any $1 \leq \alpha \leq N-1$, and after evaluating all contributions we obtain
\begin{equation} \label{eq:FluxjTpeqHomog}
\mathcal{F} = \mathcal{F}_{\alpha \to \alpha + 1} = \frac{\lambda ^2}{2 \zeta mM} \, \big( T_{\alpha} - T_{\alpha + 1} \big),
\end{equation}
where notation $\mathcal{F}_{\alpha \to \alpha + 1}$ emphasizes that heat only flows from the site $\alpha$ to its nearest-neighbor, while $\mathcal{F}$ reminds us that actually it does not depend on which site $\alpha$ we are evaluating it. This is the same result obtained in Ref.\cite{PereiraFalcao2004}. We now use Eq.\eqref{eq:FluxjTpeqHomog} to recall the next steps: since heat flux is the same all along the chain, we can add $\mathcal{F}_{\alpha \to \alpha + 1}$ for $1 \leq \alpha \leq N-1$, and noticing that we will get a telescoping sum on the right-hand side (RHS) of Eq. \eqref{eq:FluxjTpeqHomog}, we have
$$
(N-1) \, \mathcal{F} = \frac{\lambda ^2}{2 \zeta mM} \, \big( T_{1} - T_{N} \big) ,
$$
and so Fourier law holds for this model, with a thermal conductivity
\begin{equation} \label{eq:kappaTpeqHomog}
\kappa = \frac{\lambda ^2}{2 \zeta mM} \, ,
\end{equation}
that does not depend on temperature.

Now we quickly remind results for another previously studied model that is related to this first one, namely the NN ``almost" homogeneous chain, where the coupling $\zeta_{j}$ between site $j$ and its heat bath may arbitrarily change over the chain. We get
\begin{equation} \label{eq:FluxjTpeqQHomog}
\mathcal{F} = \mathcal{F}_{\alpha \to \alpha + 1} = \frac{\lambda ^2}{mM} \frac{ T_{\alpha} - T_{\alpha + 1} }{\zeta_{\alpha} + \zeta_{\alpha + 1}} ,
\end{equation}
for any $\alpha$ in the bulk of the chain, in agreement with the result obtained in Ref.\citep{PereiraFalcao2006}. Since the flux $\mathcal{F}$ must be the same all along the chain since the system is on NESS, we can mimic the previous approach to find out
\begin{equation} \label{eq:FluxTpeqQHomog}
\mathcal{F} = \frac{\lambda ^2}{mM} \left[ \sum_{1 \leq j \leq N-1} \left( \zeta_{j} + \zeta_{j+1} \right) \right]^{-1} (T_{1} - T_{N}).
\end{equation}

\subsection{Main results for NNN rotor model} \label{ssec:main_results}

We now turn to our main object of study, the rotor model. It is an almost homogeneous chain, but with NNN interaction: again we take $m_{j}=m$ and $M_{j}=M$, for any site, and we start with arbitrary heath bath-site coupling $\zeta_{j}$. Concerning anharmonic interparticle interaction, now we study the NNN model, i.e. 
\begin{equation} \label{eq:lambdaNNN}
    \lambda_{j,l} = \begin{cases}
\lambda > 0 & \mbox{, if } |j-l| = 1, \\
\nu & \mbox{, if } |j-l| = 2, \\
0 & \mbox{, otherwise}.
\end{cases}
\end{equation}
For the sake of perturbative calculations performed in \eqref{eq:perturbrot}, $\nu$ will be taken in the same order as $\lambda$, but we stress out that $\nu$ may be positive or negative, in contrast with always positive NN coupling $\lambda$. To elucidate this point, let us show an intermediary step and evaluate, for example, \eqref{eq:FtTpeq} using values for this NNN model. We get
\begin{eqnarray}
- \left\langle \Omega; F_t \right\rangle_{0} &=& \frac{\lambda^{2}}{4mM} \left[ \left( \frac{1}{\zeta_{\alpha + 1}} - \frac{2}{\zeta_{\alpha}} \right) T_{\alpha} + \left( \frac{2}{\zeta_{\alpha + 1}} - \frac{1}{\zeta_{\alpha}} \right) T_{\alpha + 1} \right] + 
\nonumber
\\
& & + \frac{\lambda \nu}{2mM} \left[ - \frac{2}{\zeta_{\alpha}} T_{\alpha} + \frac{1}{\zeta_{\alpha + 1}} T_{\alpha + 1} + \frac{1}{\zeta_{\alpha + 2}} T_{\alpha + 2} \right] + 
\nonumber
\\
& & + 
\frac{\nu^{2}}{4mM} \left[ \left( \frac{1}{\zeta_{\alpha + 2}} - \frac{2}{\zeta_{\alpha}} \right) T_{\alpha} + \left( \frac{2}{\zeta_{\alpha + 2}} - \frac{1}{\zeta_{\alpha}} \right) T_{\alpha + 2} \right].
\label{eq:FtTpeqQHomogNNN}
\end{eqnarray}
The first term on RHS of Eq. \eqref{eq:FtTpeqQHomogNNN} above is proportional to $\lambda^2$ and it is due to NN interaction only -- as we can see inside the brackets, it only depends on $\alpha$ and $\alpha + 1$. The second term is proportional to $\lambda \nu$, and it is due to $\alpha$-th site interaction both with its nearest- and next-nearest-neighbors. The third and last term is proportional to $\nu^2$, and it depends only on $\alpha$-th site interaction with his NNN, the $(\alpha + 2)$-th site.

If we take $\nu = 0$, only the first term above will be non-vanishing, and we obviously recover the NN model. However, as we turn on the NNN interaction, we could have different behaviors if $\nu$ is positive or negative. If $\nu > 0$, the second and third terms have the same sign as the first one \footnote{and its sign depends if $T_{1}>T_{N}$ or $T_{1}<T_{N}$}, so we are only increasing its contribution to thermal conductivity $\kappa$. However, if $\nu < 0$, the third term still increases $\kappa$, but the second term could decrease it. In essence, a negative NNN interaction could inhibit heat flow. Up to this point, this discussion refers only to \eqref{eq:FtTpeq} contribution to heat flow, we still must evaluate \eqref{eq:W1rot_Tpeq_calc}-\eqref{eq:W3rotorTpeq}, but we claim that this same behavior holds. In summary, after calculations, the heat flux can be written as
$$
\mathcal{F}_{\alpha \rightarrow} = \lambda^2 c_{\lambda^2} (T) + \lambda \nu c_{\lambda \nu} (T) + \nu^2 c_{\nu^2} (T) ,
$$
where those coefficients $c_{\lambda^2}$, $c_{\lambda \nu}$ and $c_{\nu^2}$ are either all positive or all negative, so again $\lambda \nu$ term could decrease the intensity of thermal conductivity for $\nu < 0$. Nevertheless, as we evaluate all contributions \eqref{eq:FtTpeq}-\eqref{eq:W3rotorTpeq} to the heat flux, we get
\begin{equation} \label{eq:fluxNNN_final}
\mathcal{F}_{\alpha \rightarrow} = \frac{\lambda ^2}{mM} \frac{ T_{\alpha} - T_{\alpha + 1} }{\zeta_{\alpha} + \zeta_{\alpha + 1}} + \frac{\nu^2}{mM} \frac{ T_{\alpha} - T_{\alpha + 2} }{\zeta_{\alpha} + \zeta_{\alpha + 2}} \, .
\end{equation}
In other words, we have $c_{\lambda \nu} = 0$, and thermal conductivity on NESS can only increase, even with a negative interaction between next-nearest neighbors. This result is in contrast with that one obtained by \cite{Olivares2016}, as they saw an insulator regime for the rotor model with LR interactions. Our result suggests that this insulator regime must be a genuine LR effect. Just as an illustration, from the equations above we obtain an expression for the heat flow in terms of the temperatures at the ends. Indeed, summing up the equations (we make $\zeta_{\alpha} = \zeta$ and consider $N$ even)

\begin{eqnarray*}
  \mathcal{F}_{1 \rightarrow} &=& \frac{\lambda ^2}{mM} \frac{ T_{1} - T_{2} }{2\zeta} + \frac{\nu^2}{mM} \frac{ T_{1} - T_{3} }{2\zeta} \\
   \mathcal{F}_{2 \rightarrow} &=& \frac{\lambda ^2}{mM} \frac{ T_{2} - T_{3} }{2\zeta} + \frac{\nu^2}{mM} \frac{ T_{2} - T_{4} }{2\zeta} \\
    \mathcal{F}_{3 \rightarrow} &=& \frac{\lambda ^2}{mM} \frac{ T_{3} - T_{4} }{2\zeta} + \frac{\nu^2}{mM} \frac{ T_{3} - T_{5} }{2\zeta} \\
     &=& \ldots \\
      \mathcal{F}_{N-2 \rightarrow} &=& \frac{\lambda ^2}{mM} \frac{ T_{N-2} - T_{N-1} }{2\zeta} + \frac{\nu^2}{mM} \frac{ T_{N-2} - T_{N} }{2\zeta} \\
      \mathcal{F}_{N-1 \rightarrow} &=& \frac{\lambda ^2}{mM} \frac{ T_{N-1} - T_{N} }{2\zeta} ,
\end{eqnarray*}
considering $ \mathcal{F}_{\alpha \rightarrow} = \mathcal{F}$, we obtain
\begin{equation*}
    (N-1)\mathcal{F} = \frac{\lambda ^2}{mM 2\zeta}(T_{1}-T_{N}) + \frac{\nu^2}{mM 2\zeta}(T_{1}-T_{N-1}) + \frac{\nu^2}{mM 2\zeta}(T_{2}-T_{N}),
\end{equation*}
that is, 
\begin{equation*}
    (N-1)\mathcal{F} \approx \frac{\lambda ^2 + 2\nu^{2}}{mM 2\zeta}(T_{1}-T_{N}) .
\end{equation*}

We aim now to investigate thermal rectification for a NNN interaction-related model. A necessary ingredient for (possible) thermal rectification is that the chain must have some asymmetry, so we set a linearly graded mass chain -- on the other hand, we simplify calculations taking $\zeta_{j} = \zeta > 0$ for all sites of the chain. If we take, without loss of generality, $m_{1} > m_{N}$, then we have $m_{j} = [(N-j) m_{1} + (j-1) m_{N}]/(N-1)$, for any $1 \leq j \leq N$. Analytical evaluations are too hard for a graded mass, so we perform numerical computations for the flux \eqref{eq:fluxTpeq}. We emphasize that we are not performing computer simulations to find dynamics evolution for this system from scratch, but rather we have first developed a perturbative analytical approach to evaluate heat flux \eqref{eq:fluxTpeq}. For a case of anharmonic interaction with inner noises and unbounded potential, a perturbative approach was rigorously proven to be convergent in previous works \cite{PereiraMendoncaLemos}. We start from this point to numerically obtain heat flux. The following ten parameters must be given as inputs: the size of the chain $N$, the heat bath-site coupling $\zeta > 0$, the on-site pinning $M>0$, both NN and NNN interactions strength, respectively $\lambda > 0$ and $\nu \in \mathbb{R}$, the factor $\kappa$, the masses $m_{1} > m_{N}$ and the temperatures $T_{1}$ and $T_{N}$ for the boundaries sites of the chain. Concerning temperatures, we remind that only temperatures at the boundaries of the chain are given, and they are labeled as $T_{H}$ and $T_{C}$, where the indices respectively stand for hot and cold baths. As previously said, remaining temperatures $T_{j}$, for any $j$ in the bulk of the chain, must be found using the self-consistency condition. And so, for each set of parameters, we find two temperature profiles: the first one for $T_{1} = T_{H}$ and $T_{N} = T_{C}$, and the other when we exchange temperatures at the boundaries. With both profiles at hand, we can evaluate the flux from left to right of the chain $\mathcal{F}_{L}$, when $T_{1} > T_{N}$, and reversed flux $\mathcal{F}_{R}$, when $T_{1} < T_{N}$. Since analytical expression \eqref{eq:fluxTpeq} was obtained considering the flux to the right, we obviously expect $\mathcal{F}_{R} < 0 < \mathcal{F}_{L}$. However, despite heat flows in opposite directions for $\mathcal{F}_{R}$ and $\mathcal{F}_{L}$, they could have the same magnitude, i.e. $|\mathcal{F}_{R}| = \mathcal{F}_{L}$, and if this is the case our model presents no thermal rectification, at least for low-temperature regime. On the other hand, if we find that $|\mathcal{F}_{R}| \neq \mathcal{F}_{L}$, we can conclude that the model is a thermal rectificator.

We used \textit{Mathics} to perform numerical calculations. Despite we have listed ten parameters as inputs in the previous paragraph above, we can change all our variables to dimensionless ones, as we show in appendix \ref{app:dimless}, and by doing so we will always have dimensionless unit values for the on-site potential $M=1.0$, the largest mass $m_{1}=1.0$, and for the NN interaction coupling $\lambda = 1.0$. In such a scenario, a low-temperature regime means that the hot thermal reservoir temperature is $T_{H} < 1.0$. So, for a small chain with $N=16$ sites, given the fixed parameters: $m_{N}=0.5$, $\zeta = 1.0$, $\kappa = 1.0$, and $\nu = -0.11 < 0$, we set hot and cold temperatures as $T_{H}=0.2$ and $T_{C}=0.1$, and our program returns a left flux $\mathcal{F}_{L} = 0.00659$ and a right flux $\mathcal{F}_{R} = -0.00215$, and so we have a thermal rectification. If we change NNN-coupling to a positive value $\nu = 0.11 > 0$, keeping all the other parameters at their same values, we get the same fluxes $\mathcal{F}_{L} = 0.00659$ and $\mathcal{F}_{R} = -0.00215$, and this was expected, since the heat flux \eqref{eq:fluxNNN_final} only depends on $\nu^{2}$. If we double chain size to $N=32$, we roughly obtain half the fluxes, i.e. $\mathcal{F}_{L} = 0.00328$ and $\mathcal{F}_{R} = -0.00104$, which suggests that the flux decays on chain size $N$, but our chains are too small to conjecture any conclusion on such dependence. In the table below we list some values for the parameters and the fluxes.

\begin{center}
\begin{tabular}{|c|c|c|c|c|c|}
    \hline
    $N$                   & $T_{H}$ & $T_{C}$ & $\mathcal{F}_{L}$ & $\mathcal{F}_{R}$ & $\mathcal{F}_{L} + \mathcal{F}_{R}$  \\
    \hline
    \multirow{4}{4em}{16} & 0.2      & 0.1     & 0.0065949        & -0.00214865        & 0.00444625                          \\ 
                          & 0.3      & 0.1     & 0.0117077        & -0.00577938        & 0.00592834                          \\ 
                          & 0.5      & 0.1     & 0.0219334        & -0.0130409         & 0.00889251                          \\
                          & 0.4      & 0.2     & 0.0131898        & -0.0042973         & 0.00889251                          \\ 
    \hline
    \multirow{4}{4em}{32} & 0.2      & 0.1     & 0.00321816       & -0.00103947        & 0.00217869                          \\ 
                          & 0.3      & 0.1     & 0.0571009        & -0.00280516        & 0.00290493                          \\ 
                          & 0.5      & 0.1     & 0.0106939        & -0.0633656         & 0.00435739                          \\
                          & 0.4      & 0.2     & 0.0643632        & -0.00207893        & 0.00435739                          \\ 
    \hline
\end{tabular}
\end{center}

\section{Final Remarks} \label{sec:conclusion}

In this article, we investigate the heat flow and rectification in a version of the rotor model (here, with inner stochastic noises), a version involving interactions with next-nearest neighbors.  It is worth recalling that the original rotor chain, in the case of long-range interactions, presents an interesting behavior, precisely, the existence of an insulating regime. Hence, our main interest was to find some hint of such an insulating regime, that is, a possible decay in the heat flow with the addition of the next-neighbor interaction to the nearest-neighbor one. We show, however, in a perturbative analytical computation, that such an addition of the next-neighbor interaction increases the heat flow even so the sign of the coupling interaction is negative (or positive). Our result indicates that the insulating regime is characteristic of real long-range interaction, that is, at least up to the next-neighbor case, it seems to be absent for short-range interaction. We reinforce that the detailed analytical study of models recurrently investigated by numerical methods is important and may help us to better understand what is happening. 

Besides the investigation of homogeneous chains, we consider graded asymmetric systems by performing numerical computations. In this case, we show the occurrence of thermal rectification.

\begin{acknowledgments}
Some acknowledgements.
\end{acknowledgments}

\bibliography{biblio}

\appendix

\section{Dimensionless units} \label{app:dimless}

After analytically evaluating the expression for the heat flow for the rotor model -- see Eqs.\eqref{eq:FtTpeq}, \eqref{eq:W1rot_Tpeq_calc}, \eqref{eq:W2rotTpeq} e \eqref{eq:W3rotorTpeq} -- and show that, at least in low-temperature regime, the heat flux only increases after we introduce a NNN coupling, we decided to study them numerically aiming to investigate if this model rectifies heat flux. However we have a large number of parameters, so we performed a dimensionless study for the model, aiming to decrease the number of free parameters involved. Moreover, such dimensionless study allows us to precisely define the low-temperature regime. Despite we have started with a Hamiltonian with general onsite and interparticle interactions \eqref{eq:H}, here, for the sake of understanding, we rewrite the specific Hamiltonian for our model studied in subsection \ref{ssec:main_results}
    \begin{equation} \label{eq:rot_H}
        \mathcal{H} = \sum _{j=1}^N \left( \frac{p_j^2}{2m_j} + \frac{1}{2} \, m \omega^{2} q_{j}^{2} \right) + \lambda \sum _{j=1}^{N-1} [1 - \cos (\kappa (q_{j}-q_{j+1}))] + \nu \sum _{j=1}^{N-2} [1 - \cos (\kappa (q_{j}-q_{j+2}))]  ,
    \end{equation}
where we have already used that we study the case where all onsite harmonic potential have the same strength $M_{j}=M$, we used this to define a natural frequency for the system as $\omega^{2} = M/m$. However, since the model has different graded masses $m_{j}$, we have chosen $m=m_{1}$ as the largest one. We also rewrite the dynamics for this specific model as
\begin{subequations} \label{eq:rot_dyn}
\begin{align}
\label{eq:rot_dynq}
dq_j =& \frac{\partial \mathcal{H}}{\partial p_j} \, dt = \frac{p_j}{m_j} \, dt ,
\\
dp_j =& - m \omega^{2} q_j dt - \lambda \kappa [ \sin (\kappa (q_{j} - q_{j-1})) + \sin (\kappa (q_{j} - q_{j+1})) ] \, dt + \nonumber \\
& + (-\mu \kappa) [ \sin (\kappa (q_{j} - q_{j-2})) + \sin (\kappa (q_{j} - q_{j+2})) ] \, dt - \zeta  p_j dt + \gamma _j^{1/2} dB_j . \label{eq:rot_dynp}
\end{align}
\end{subequations}
Surely, we must be careful with the $dp_{j}$ equations for the sites at the boundaries of the chain, namely for $j=1$, $j=2$, $j=N-1$, or $j=N$. But the main idea here is just to understand our system using dimensionless units, and those equations for the dynamics of the boundaries of the chain can be treated similarly.

Concerning the parameters: $q_{j}$ and $p_{j}$ are respectively position and momentum for the $j$-th particle of the chain, and so they have their usual units; $m$ and $m_{j}$ are masses; $\omega$ is a frequency, and so $M = m \omega^{2}$ has force per length unit. We can see from Hamiltonian \eqref{eq:rot_H} that both $\lambda$ and $\mu$ have energy units. From dynamics \eqref{eq:rot_dynp}, we can see that $\zeta$ has frequency units; the temperatures $T_{j}$ have energy units, and $dB{j}$ has the unit of the square root of time -- a well-known fact from stochastic calculus that can be seen on Eq. \eqref{eq:dB}.

We start re-scaling the energy of the system to a dimensionless Hamiltonian $\hat{\mathcal{H}}$ defined by $\mathcal{H} = \lambda \hat{\mathcal{H}}$, so
$$
    \hat{\mathcal{H}} = \sum _{j=1}^N \left( \frac{1}{2} \, \frac{p_j^2}{\lambda m_j} + \frac{1}{2} \, \frac{m \omega^{2}}{\lambda} q_{j}^{2} \right) + \sum _{j=1}^{N-1} [1 - \cos (\kappa (q_{j}-q_{j+1}))] + \frac{\nu}{\lambda} \sum _{j=1}^{N-2} [1 - \cos (\kappa (q_{j}-q_{j+2}))] .
$$
We further comment on this choice at the end of this appendix. We can notice that now we can define a dimensionless NNN interaction $\hat{\nu} = \nu/ \lambda$, but most important: we can see from the equation above that we have re-scaled the NN interaction strength to the unit, i.e. $\hat{\lambda} \equiv 1$, and so the NN interaction strength is our first parameter that is fixed to the unit when we use dimensionless units. We can also define dimensionless position $\hat{q}_{j}$ as
\begin{equation} \label{eq:dimless_qj}
    \hat{q}_{j} = \sqrt{\frac{m \omega^{2}}{\lambda}} \, q_{j} ,
\end{equation}
and that is the same to consider that we have unit frequency, or equivalently that $\hat{M} \equiv 1$; in other words, the on-site potential $M$ is our second parameter fixed to the unit. We could define right now the dimensionless momentum, but we would rather keep clear that we have different graded masses along the chain as we define dimensionless masses
\begin{equation} \label{eq:dimless_mj}
    \hat{m}_{j} = \frac{m_{j}}{m} = \frac{m_{j}}{m_{1}} \, .
\end{equation}
It is clear from the definition above that $\hat{m}_{1} \equiv 1$, therefore the mass of the first particle of the chain $m_{1}$ is our third and last parameter fixed to the unit. And now we define dimensionless momentum $\hat{p}_{j}$ as
\begin{equation} \label{eq:dimless_pj}
    \hat{p}_{j} = \frac{1}{\sqrt{m \lambda}} \, p_{j} ,
\end{equation}
and we are almost set to write our dimensionless Hamiltonian. The last missing piece is to keep sine function arguments dimensionless, defining $\hat{\kappa}$ as
$$
\hat{\kappa} \hat{q}_{j} = \kappa q_{j} \Rightarrow \hat{\kappa} = \sqrt{\frac{\lambda}{m \omega^{2}}} \, \kappa.
$$
We finally get
\begin{equation} \label{eq:dimless_H}
    \hat{\mathcal{H}} = \sum _{j=1}^N \left( \frac{\hat{p}_j^2}{2 \hat{m}_j} + \frac{1}{2} \, \hat{q}_{j}^{2} \right) + \sum _{j=1}^{N-1} [1 - \cos (\hat{\kappa} (\hat{q}_{j}-\hat{q}_{j+1}))] + \hat{\nu} \sum _{j=1}^{N-2} [1 - \cos (\hat{\kappa} (\hat{q}_{j}-\hat{q}_{j+2}))] .
\end{equation}

Now we will check dynamics to be dimensionless. Our first guess to a dimensionless time would be $\tau = \omega t$, and indeed it is the right choice. To properly do so, let us check consistency for the first equation of the dynamics \eqref{eq:rot_dynq}. For dimensionless position, we get
\begin{eqnarray}
    d \hat{q}_{j} &=& d \left( \sqrt{\frac{m \omega^{2}}{\lambda}} \, q_{j} \right) = \sqrt{\frac{m \omega^{2}}{\lambda}} \, dq_{j} = \sqrt{\frac{m}{\lambda}} \, \omega \frac{p_{j}}{m_{j}} dt = \nonumber \\
    &=& \sqrt{\frac{m}{\lambda}} \, \frac{\sqrt{m \lambda} \hat{p}_{j}}{m \hat{m}_{j}} \omega dt = \frac{\hat{p}_{j}}{\hat{m}_{j}} d \tau = \frac{\partial \hat{\mathcal{H}}}{\partial \hat{p}_{j}} \, d \tau, \label{eq:dimless_dynq}
\end{eqnarray}
where we have defined $d \tau = \omega dt$ to keep dynamics consistent, therefore $\tau = \omega t$ really is our dimensionless time. To deal with the remaining parameters, we turn to the dimensionless version for equation \eqref{eq:rot_dynp}, which is
\begin{eqnarray*}
    d \hat{p}_{j} &=& d \left( \frac{1}{\sqrt{m \lambda}} \, p_{j} \right) = \frac{1}{\sqrt{m \lambda}} \, dp_{j} = \\
    &=&-\frac{m \omega^{2}}{\sqrt{m \lambda}} \, \sqrt{\frac{\lambda}{m \omega^{2}}} \, \hat{q}_{j} \, \frac{1}{\omega} \, d \tau - \frac{\lambda}{\sqrt{m \lambda}} \sqrt{\frac{m \omega^{2}}{\lambda}} \, \hat{\kappa} [ \sin (\hat{\kappa} (\hat{q}_{j} - \hat{q}_{j-1})) + \sin (\hat{\kappa} (\hat{q}_{j} - \hat{q}_{j+1})) ] \frac{1}{\omega} \, d \tau + \\
    & & + \left( - \frac{\nu}{\sqrt{m \lambda}} \right) \sqrt{\frac{m \omega^{2}}{\lambda}} \, \hat{\kappa} [ \sin (\hat{\kappa} (\hat{q}_{j} - \hat{q}_{j-1})) + \sin (\hat{\kappa} (\hat{q}_{j} - \hat{q}_{j+1})) ] \frac{1}{\omega} \, d \tau + \\
    & & + \left( - \frac{\zeta}{\sqrt{m \lambda}} \right) \sqrt{m \lambda} \, \hat{p_{j}} \, \frac{1}{\omega} \, d \tau + \frac{\sqrt{2 \zeta m_{j} T_{j}}}{\sqrt{m \lambda}} \,  dB_{j} ,    
\end{eqnarray*}
and after some calculations, we get
\begin{eqnarray}
    d \hat{p}_{j} &=& - \hat{q}_{j} d \tau - \hat{\kappa} [ \sin (\hat{\kappa} (\hat{q}_{j} - \hat{q}_{j-1})) + \sin (\hat{\kappa} (\hat{q}_{j} - \hat{q}_{j+1})) ] d \tau + \nonumber \\
    & & + (-\hat{\nu} \hat{\kappa}) [ \sin (\hat{\kappa} (\hat{q}_{j} - \hat{q}_{j-1})) + \sin (\hat{\kappa} (\hat{q}_{j} - \hat{q}_{j+1})) ]  d \tau - \frac{\zeta}{\omega} \, \hat{p}_{j} d \tau + \sqrt{\frac{2 \zeta \hat{m}_{j} T_{j}}{\lambda}} \, dB_{j} =  \nonumber\\
    &=& - \frac{\partial \hat{\mathcal{H}}}{\partial \hat{q}_{j}} \, d \tau - \frac{\zeta}{\omega} \, \hat{p}_{j} d \tau + \sqrt{\frac{2 \zeta \hat{m}_{j} T_{j}}{\lambda}} \, dB_{j}. \label{eq:dimless_dynp_temp}
\end{eqnarray}
The expression above shows us that the Hamiltonian part of the dynamics for $d \hat{p}_{j}$ is consistent, we only must finish it by defining our last dimensionless quantities. We can now define the dimensionless coupling between the $j$-th site of the chain and its thermal reservoir, $\hat{\zeta} = \zeta / \omega$. We also have dimensionless temperature $\hat{T}_{j} = T_{j} / \lambda$, and now we can discuss more precisely what is the low-temperature regime: in this dimensionless study, as $\hat{\lambda} = 1$, we must have $\hat{T}_{j} < 1$ for any site $j$ of the chain. This can be obtained if we set our hot thermal reservoir $\hat{T}_{H} < 1$.
Finally, debunking our last term on \eqref{eq:dimless_dynp_temp}, we have
$$
\sqrt{\frac{2 \zeta \hat{m}_{j} T_{j}}{\lambda}} \, dB_{j} = \sqrt{2 \omega \hat{\zeta} \hat{m}_{j} \hat{T}_{j}} \, dB_{j} = \sqrt{2 \hat{\zeta} \hat{m}_{j} \hat{T}_{j}} \, d\hat{B}_{j},
$$
where we have defined dimensionless Brownian motion $d\hat{B}_{j} = \omega^{\frac{1}{2}} dB_{j}$. We first remind that $dB_{j}$ has the same unit of $dt^{\frac{1}{2}}$, so our definition is clearly dimensionless. Moreover, from \eqref{eq:dB}, we have
$$
\langle d\hat{B}_{j} (\tau) d \hat{B}_{j'} (\tau ') \rangle = \omega \langle dB_{j} (t) dB_{j'} (t') \rangle = \delta_{j,j'} \delta (t-t') \omega dt = \delta_{j,j'} \delta (\tau - \tau ') d \tau.
$$
So, we conclude the dimensionless dynamics
\begin{equation}
    d \hat{p}_{j} =- \frac{\partial \hat{\mathcal{H}}}{\partial \hat{q}_{j}} \, d \tau - \hat{\zeta} \hat{p}_{j} d \tau + \sqrt{2 \hat{\zeta} \hat{m}_{j} \hat{T}_{j}} \, d\hat{B}_{j}.
\end{equation}

As a final comment, we discuss about our choice to use $\lambda$ to re-scale to a dimensionless Hamiltonian. As it was told before, such a choice fixes our dimensionless NN interaction strength to $\hat{\lambda} = 1$, and so it should not be used to perform a full study of our model. We first remind that we started solving a simpler stochastic process, and to do so we have set $\lambda = 0$ -- see Eq. \eqref{eq:dynphase0}. Moreover, after using the Girsanov theorem to recover NN and NNN interparticle interactions and construct an integral formalism, we used $\lambda$ as a small parameter for a perturbative study, see Eq. \eqref{eq:perturbrot}. Both reasons make $\lambda$ unfeasible to be used as a re-scaling parameter. However, our goals in this section were only to reduce the number of free parameters of our model, aiming to perform some numerical calculations, and incidentally to understand what is the low-temperature regime. For both these intuit $\lambda$ can be used as our re-scailing parameter for the Hamiltonian.

\end{document}